\documentclass[a4paper,11pt]{article}
\pdfoutput=1 % if your are submitting a pdflatex (i.e. if you have
\usepackage{jheppub} % for details on the use of the package, please
\usepackage{amsmath}
\allowdisplaybreaks[4]        
\usepackage{amssymb}
\usepackage{euscript}     
\usepackage{color}         
\usepackage{tensor}        
\usepackage{amsthm}

\usepackage[header,title,page,titletoc]{appendix}  %010078010061010087010071appendi
\usepackage[T1]{fontenc} % if needed
\usepackage[numbers]{natbib}  
\def\ll{ c}
\def\z{{\bar{z}}}
\renewcommand{\(}{\left(}
\renewcommand{\)}{\right)}

\title{\boldmath Horizon supertranslation and degenerate black hole solutions}

\author[a,b]{Rong-Gen Cai}
\author[a]{Shan-Ming Ruan}
\author[c]{Yun-Long Zhang}

\affiliation[a]{CAS Key Laboratory of Theoretical Physics, Institute of Theoretical Physics,\\ Chinese Academy of Sciences, Beijing 100190, China}
\affiliation[b]{Center for Gravitational Physics, Yukawa Institute for Theoretical Physics,\\ Kyoto University, Kyoto 606-8502, Japan }
\affiliation[c]{Asia Pacific Center for Theoretical Physics,\\Pohang 790-784, Korea}

\emailAdd{cairg@itp.ac.cn}
\emailAdd{ruanshanming@itp.ac.cn}
\emailAdd{yunlong.zhang@apctp.org}

\abstract{In this note we  first review the degenerate vacua arising from the BMS symmetries. According to the discussion in~\cite{Donnay:2015abr} one can define BMS-analogous supertranslation and  superrotation for spacetime with black hole in Gaussian null coordinates. In the leading and subleading orders of near horizon approximation, 
	the infinitely degenerate black hole solutions are derived by considering Einstein equations with or without cosmological constant, and they are related to each other by the diffeomorphism generated by horizon supertranslation. Higher order results and degenerate Rindler horizon solutions also are given in appendices.  }

\begin{document} 
\maketitle
\flushbottom
	\section {Introduction}
	Almost half a century ago, Bondi, van der Burg, Metzner and Sachs (BMS) \cite{Bondi:1962px,Sachs:1962wk}   independently investigated gravitational waves near the null infinity in asymptotically flat spacetime and showed that the spacetime has an  infinitesimal dimensional group associated with the asymptotic symmetries called BMS group now. A few years later, Weinberg found that  there is a universal soft theorem \cite{Weinberg:1965nx,Weinberg:1995mt} relating  one $\mathcal{S}$-matrix element of n-particles to the other with an additional  zero four-momentum photon or graviton that is generally called soft particle, which plays an important role in eliminating the infrared divergence in quantum field theory. In recent years, Strominger got some insights on the infrared structure of quantum gravity \cite{Strominger:2013jfa} and connected these two  seemingly different  matters mentioned above with his collaborators. It is verified that the soft graviton theorem is exactly equivalent to the Ward identity of the BMS supertranslation \cite{He:2014laa,Kapec:2015vwa} and there is also equivalence between the subleading soft theorem \cite{Cachazo:2014fwa} and the Ward identity of superotation \cite{Kapec:2014opa,Adamo:2014yya,Geyer:2014lca}. Furthermore,  the soft theorems and asymptotic symmetries are related to the traditional gravitational memory effect \cite{Strominger:2014pwa} and new spin memory effect \cite{Pasterski:2015tva}. In addition, these elegant connections and equivalences are also found in gauge theories. The large gauge symmetries \cite{Strominger:2013lka} of gauge theories act as the asymptotic symmetries at  the null infinity just like  in  the BMS group. Of course, BMS transformation can also be considered as large diffeomorphism. Ward identity of large gauge transformation is found to be equivalent to  the soft photon \cite{He:2014cra,Kapec:2015ena,Campiglia:2015qka,Kapec:2014zla,Mohd:2014oja} or gluon theorems \cite{He:2015zea} which are related to the observable effect, i.e., electromagnetic memory \cite{Pasterski:2015zua}. So all these series of works, starting from  fifty years ago, finally illustrate the wonderful triangular connections among soft factors, symmetries and memories for gauge field theory and gravity theory \cite{Pasterski:2015zua,Susskind:2015hpa}. The equivalence is also extended to fermionic symmetry in \cite{Dumitrescu:2015fej,Avery:2015iix,Lysov:2015jrs} where the authors showed that the Ward identity of residual local supersymmetry can be understood as soft gravitino theorem.

	The asymptotic symmetries or large gauge symmetries not only  are related to the BMS group and soft theorems, but also stimulate new insight about black hole physics. Hawking, Perry and Strominger (HPS) \cite{Hawking:2016msc} recently noticed that there is an infinite family of degenerate vacua, associated with an asymptotically flat spacetime,
	because of BMS supertranslation which enables black hole to carry soft hair storing information about matter. At the same time, black hole must also carry soft gauge hairs due to the infinite conservation laws coming from large abelian gauge symmetries if Maxwell field is present in the theory under consideration. This claim points out the  flaws of Hawking's original argument about the information loss paradox~\cite{Hawking:1974sw,Hawking:2016msc}. Although the HPS's proposal has not yet solved the information loss paradox, these infinite soft hairs of black hole actually indicate a hopeful  direction for information problem of black hole. In addition to the information loss paradox, degenerate black hole states also  enable us to count the microstates of black hole \cite{Hotta:2016qtv,Averin:2016ybl} although it is not exact in the four-dimensional case  due to the absence of fully understanding about superrotation. But, some works \cite{Afshar:2016wfy,Afshar:2016uax,Sheikh-Jabbari:2016npa} give very nice results about the  microstate counting  in the case of 3-dimensional black hole.

	Note that  most of recent works relevant to  the HPS's proposal focus on the symmetries at the null infinity.  We here want to directly investigate the asymptotic symmetries near the horizon inspired by the works in \cite{Donnay:2015abr,Donnay:2016ejv} where the authors showed that there are BMS-analogous supertranslation and superrotation at the horizon. In this note, we first review the BMS supertranslation and degenerate vacuum solutions, then discuss the horizon supertranslation and  find the corresponding infinite degenerate black hole solutions of Einstein equations with or without cosmological constant in the near horizon regime.

	\section{BMS supertranslation and degenerate vacua}

	In this section, we simply review the basics  of  the supertranslation in BMS group \cite{Barnich:2009se,Barnich:2010eb,Barnich:2011mi} and the resulting infinitesimal degenerate vacuum states which play a pivot role in understanding the soft hair of black hole. Let us start from the general BMS metric ansatz that can represent asymptotically flat spacetimes\footnote{Here we follow the notions in \cite{Barnich:2009se,Barnich:2011mi}. The PHD thesis \cite{Lambert:2014poa} is a simple and good review for BMS gauge in four and three dimensions.}
	\begin{equation}
	ds^2=e^{2\beta}\frac{V}{r}du^2 -2e^{2\beta}dudr +g_{AB}(dx^A-U^Adu)(dx^B-U^Bdu),
	\end{equation}	
	with  four gauge fixing conditions
	\begin{equation}\label{BMS gauge}
	g_{rr=0}, \qquad g_{rA}=0, \qquad {\rm det}(g_{AB})=r^4{\rm det}(\gamma_{AB}),
	\end{equation}
	where	$\gamma_{AB}$ is the metric of two-dimensional sphere whose coordinates are described by indices $A,B$ and associated covariant derivative is $\bar{D}$. 
	By requiring the Lie derivative $\mathcal{L}_{\zeta}g_{\mu\nu}$ generated by vector $\zeta$  defining the asymptotic symmetries to satisfy gauge fixing conditions (\ref{BMS gauge}) and some fall-off boundary conditions, we can find the solutions of the vector as
	\begin{equation}
	\zeta=( T+\frac{u}{2}\bar{D}_CR^C) \partial_u -\frac{r}{2}(\bar{D}_A\zeta^A -U^C\partial_C f)\partial_r+(R^A-\partial_B T\int^{\infty}_{r'}e^{2\beta}g^{AB}dr')\partial_A,
	\end{equation}
	where $T$ and $R$ are functions of $x^A$ and define the supertranslation and conformal transformations, respectively.   If one allows $R(x^A)$ to have pole singularities \cite{Barnich:2009se,Banks:2003vp}, the global conformal transformation can be extended to be a local one which is generally called superrotation. But there exist some debates \cite{Lambert:2014poa,Flanagan:2015pxa} on whether the superrotation or extended BMS group is physical or not. We will  therefore consider supertranslation only in what follows. The asymptotic Killing vector $\zeta$ is derived off shell, which means that it does not rely on equations of motion.  Once the Einstein equations are imposed, the on-shell retarded Bondi coordinates can be simply expressed as\footnote{Here we adopt the simplified notions in \cite{Strominger:2013jfa,Kapec:2014opa} which are convenient for discussions on soft theorem and soft hair. } 
	\begin{equation}\label{Bondi coordinates}
	\begin{split}
	ds^2=&-du^2-2dudr +2r^2\gamma_{z\bar{z}}dzd\bar{z} \\
	&+\frac{2m_B}{r}du^2+rC_{zz}dz^2+rC_{\bar{z}\bar{z}}d\bar{z}^2+D^zC_{zz}dudz+D^{\bar{z}}C_{\bar{z}\bar{z}}dud\bar{z}\\
	&+\frac{1}{4r^2}C_{zz}C^{zz}dudr+\gamma_{z\bar{z}}C_{zz}C^{zz}dzd\bar{z}+... ~,
	\end{split}
	\end{equation}  
	where $\gamma_{z\bar{z}}=2/(1+z\bar{z})^2$ is the metric of the two-dimensional sphere defined on complex stereographic coordinate $z=e^{i\phi}\cot(\theta/2)$ and the ellipsis stands for  higher order terms which do not affect the following constraint equations. In general,  $m_B(u,z,\bar{z})$ is called Bondi mass  aspect constrained by  the component of the Einstein equations
	\begin{equation}\label{constraint equation}
	\lim_{r\rightarrow \infty}r^2G_{uu} = -2\partial_u m_B -\frac{1}{2}N_{zz}N^{zz} +\frac{1}{2}D^zD^zN_{zz} +\frac{1}{2}D^{\bar{z}}D^{\bar{z}}N_{\bar{z}\bar{z}}=	\lim_{r\rightarrow \infty}8\pi Gr^2T_{uu},
	\end{equation} 
	where $N_{zz}=\partial_u C_{zz}$ is the Bondi news tensor, which is relevant to the gravitational radiation and $D_A$ stands for the covariant derivative associated with the metric $\gamma_{z\bar{z}}$.
	More precisely, one can get \cite{Lambert:2014poa,Barnich:2010eb} 
	\begin{equation}
	\begin{split}
	g_{uz} &=\frac{1}{2}D^zC_{zz} +\frac{2}{3r}N_z +\frac{1}{6r}C_{zz}D_zC^{zz} +O(\frac{1}{r^2}),\\
	g_{AB} &=r^2\gamma_{AB}+rC_{AB} + \frac{1}{2}\gamma_{AB}C^{zz}C_{zz} +\frac{E_{AB}}{r}+O(\frac{1}{r^2}),
	\end{split}	
	\end{equation}	
	where $N_A$ is angular momentum aspect and $C_{z\bar{z}}=0$, $ E_{z\bar{z}}=0$ are determined by determinant condition in (\ref{BMS gauge}).  We only consider the supertranslation generated by the vector \cite{Barnich:2011mi,He:2014laa} 
	\begin{equation}
	\zeta=T\partial_u -\frac{1}{r}(D^zT\partial_z +D^{\bar{z}}T\partial_{\bar{z}}) +(D^zD_z T)\partial_r,
	\end{equation}
	whose surface charge can be defined as
	\begin{equation}
	Q_T =\frac{1}{4\pi G}\int_{\mathcal{I}^+_-}d^2z \gamma_{z\bar{z}}T(z,\bar{z})m_B \qquad\text{with} \qquad \{Q_T,Q_{T'}\}=0 ,
	\end{equation}
	where  $\mathcal{I}^+_-$ represents the past boundary of future null infinity $\mathcal{I}^+$. It can generate infinitesimal supertranslation transformation like
	\begin{equation}
	\begin{split}
	\mathcal{L}_{\zeta}m_B &= T\partial_um_B,\\
	\mathcal{L}_{\zeta}C_{zz}&=TN_{zz}-2D_zD_zT,
	\end{split}
	\end{equation}   
	which will transform one solution of (\ref{Bondi coordinates}) into another one. We can define the vacuum state\footnote{See the discussion about the Christodoulou-Klainerman space in   \cite{Strominger:2013jfa} and recent study about vacua of gravitational field in \cite{Compere:2016jwb} } as $N_{zz}=0$ which means there is no gravitational radiation. So the vacuum can be labelled by the $u$-independent function $C_{zz}=-2D_zD_zC(z,\bar{z})$ that obeys the boundary condition. Amazingly, BMS supertranslation teaches us a lesson that the vacuum is not unique and these infinite degenerate vacua are physically distinct and are related to each other by the BMS supertranslation which leads to a change like  $C\rightarrow C+T(z,\bar{z})$ \cite {Strominger:2014pwa}. On the other hand, non-constant BMS supertranslation will be spontaneously broken, which creates soft graviton viewed as Goldstone boson. Furthermore, gravitational memory effect \cite{Strominger:2014pwa,Pasterski:2015tva} makes us able to measure the transition of spacetime metric that is induced by radiation through null infinity. HPS \cite{Hawking:2016msc} proposed  that BMS symmetries and large gauge symmetries in abelian gauge theory enable black hole to carry soft supertranslation hair and soft electric hair, which can carry some information and shed light on resolution of the information loss paradox. But in the Bondi coordinates (\ref{Bondi coordinates}) it is not direct to find whether there is a black hole in the bulk because this kind of coordinates is designed to pay attention on the null infinity-the boundary of the spacetime. So it should be  more convenient to analyze  the analogous asymptotic symmetries of BMS group directly on the horizon. In this aspect, HPS also made  some fundamental and important discussions about the horizon supertranslation in their paper.  We expect that the  horizon supertranslation should be able to repeat similar results mentioned above and the conclusions about BMS supertranslation at the null infinity,  inspired by the work in \cite{Donnay:2015abr} and their recent extended work \cite{Donnay:2016ejv} where the authors showed that event horizon also exhibits the  familiar asymptotic symmetries generated by supertranslation and superrotation.  With the presence of Maxwell field, the authors in \cite{Mao:2016pwq} also show that isolated horizon carries a large amount of soft electric hairs which can be considered as the counterpart of soft electric hair discussed in HPS's paper \cite{Hawking:2016msc}.

	\section {Horizon supertranslation and degenerate black hole}

	Null infinity $\mathcal{I}$ can be considered as the boundary of an asymptotically  spacetime. On the other hand, black hole horizon shares some similarities with the null infinity and can be understood as another boundary of  the spacetime outside the  black hole. So it is reasonable to generalize the discussions near the null infinity to the horizon. 
	In this section we want to study the supertranslation on a black hole horizon. From the previous works, we know the fact that even after one chooses some gauge conditions generally called coordinate conditions in gravity theory, in order to eliminate the extra degrees of freedom, we still have the residual gauge transformations which are called large gauge transformation or large diffeomorphism for gravity. Actually horizon supertranslations have been studied many years ago with other motivations \cite{Hotta:2000gx,Koga:2001vq}. Of course inspired by HPS's work  \cite{Hawking:2016msc}, there have been also other works \cite{Donnay:2015abr,Averin:2016ybl,Averin:2016hhm,Hotta:2016qtv,Donnay:2016ejv} concerning supertranslation or superrotation on a black hole horizon recently. Here we want to describe how the  degenerate black hole spacetime near the horizon can appear when we have fixed the gauge and we use these degenerate BH solutions to discuss the black hole's ability to store information about the initial state. At the end of this section, we will discuss a little about gravitational memory effect near the horizon which can be considered as the method to measure the information of black hole. 
	Let's start from the simplest Schwarzschild black hole written in the infalling Eddington-Filkenstein coordinates
	\begin{equation}\label{Schwarzschild}
	ds^2 = - \(1-\frac{2m}{r}\)dv^2 + 2dv dr +2 r^2\gamma\indices{_{z\bar{z}}}dzd\bar{z}.
	\end{equation}
	We label this kind of special metric as ${g}_{0\mu\nu}$ to distinguish it from other degenerate black hole solutions. If one wants to use the near horizon geometry to discuss the infinite dimensional symmetries near the horizon just like the BMS group near the null infinity, the deviation from the metric of black hole (\ref{Schwarzschild}) should be the order of $(r-r_h)^n$ with $n > 0$. It will be convenient to introduce  the new radial coordinate $\rho=r-r_h$ to describe the region near the horizon located at $r_h$. In this coordinate,  the Schwarzschild metric can be expressed as 
	\begin{equation}\label{Schwarzschild_2}
	ds^2 = \left(-\frac{\rho}{r_h}+\frac{\rho^2}{r_h^2} \,\right)dv^2 + 2dvd\rho +2(\rho +r_h)^2 \gamma\indices{_{z\bar{z}}}dzd\bar{z},
	\end{equation}  
	with some neglected higher order terms of $\rho$. Obviously, it is not the universal near horizon geometry. On the one hand, we need to fix some gauge conditions in order to find the large diffeomorphism. On the other hand, we also need to choose suitable boundary conditions to describe the physical process near the horizon. With the motivation to define the  supertranslation on the black hole horizon, we take the four gauges as~\footnote{In the inverse metric form the coordinate conditions can be written as $g\indices{^{vv}}=0, g\indices{^{v A}}=0, g\indices{^{\rho v}}=1$. } 
	\begin{equation}\label{gauge condition}
	g\indices{_{\rho\rho}}=0, \ \ g\indices{_{\rho A}}=0, \ \ g\indices{_{\rho v}}=1,
	\end{equation}   
	which are the same as those in~\cite{Donnay:2015abr,Donnay:2016ejv,Averin:2016ybl} but a little different from the BMS gauges due to the difference in the fourth condition. We will discuss the difference at the end of this section. Actually, the coordinates satisfying these special coordinate conditions (\ref{gauge condition}) are commonly known as Gaussian null coordinates (GNC) as the analogues of Gaussian
	normal coordinates. An arbitrary null surface can be rewritten in GNC \cite{Parattu:2015gga,Chakraborty:2015aja}, so the gauge conditions are universal for any isolated horizon. For a simple example, a general black hole solution can be expressed as 
	\begin{equation}
	ds^2= -f(r)dt^2 +\frac{dr^2}{f(r)} +g_{AB}dx^Adx^B,
	\end{equation}
	and it can be  easily rewritten as 
	\begin{equation}
	ds^2= -f(r)dv^2 +2drdv +g_{AB}dx^Adx^B, \quad \text{with} \qquad 
	v=t+\int \frac{dr}{f(r)}.
	\end{equation}
	Assuming the event horizon located at $r=r_h$, one can find $f(r) \approx 2\kappa (r-r_h) $.        
	From \cite{Averin:2016ybl} we can find that this kind of gauge conditions play an important role in defining supertranslation at the horizon. For other fall off conditions for other components of metric, we follow  Ref. \cite{Donnay:2015abr} where they showed that the asymptotic symmetries  near the  horizon of black hole  are generated by charges of supertranslation and Virasoro algebra~\footnote{Recently this kind of conditions are also extended to a more general case which admits dependence of time \cite{Donnay:2016ejv}. }, and they can be expressed explicitly as 
	\begin{equation} \label{asymptotic conditions}
	\begin{split}
	g_{vv} &=-2\kappa \rho + \rho^2 \tilde{a}^{(2)}(v,z,\bar{z}) +O(\rho^{2+\epsilon}),  \\
	g_{vA} &= \rho \theta_A(v,z,\bar{z})+\rho^2 \theta^{(2)}_A(v,z,\bar{z}) +O(\rho^{2+\epsilon}), \\
	g_{AB} &= \Omega(z,\bar{z}) \gamma_{AB} + \rho \tilde{\lambda}_{AB}(v,z,\bar{z})+ \rho^2\tilde{\lambda}^{(2)}_{AB}(v,z,\bar{z})   +O(\rho^{2+\epsilon}) ,
	\end{split}
	\end{equation}
	where $A,B$ are the complex coordinate $z$ and $\bar z$ indices on the unit 2-dimensional sphere and $O(\rho^{2+\epsilon}) $ represents the higher order terms which are irrelevant in our following discussions. It can be shown that general stationary black hole can be written in this kind of form with different surface gravity $\kappa$ and function $\Omega$.  For rotating (Kerr) black hole, please see \cite{Booth:2012xm} for more detailed coordinate transformation.
	Of course, the Schwarzschild black hole satisfies the same asymptotic conditions by defining $\kappa=1/2r_h$ and $\Omega=r_h^2$. So the metric ansatz  near the  horizon takes the form as 
	\begin{equation}\label{metric}
	\begin{split}
	ds^2  & = \left(-2\kappa \rho+(\frac{\rho}{r_h})^2 \right)dv^2 + 2dvd\rho +2(\rho +r_h)^2 \gamma\indices{_{z\bar{z}}}dzd\bar{z} \\
	&  +\rho^2a^{(2)} dv^2+ 2 \rho \theta_z dzdv + 2\rho \theta_{\bar{z}}  d\bar{z}dv +\rho \lambda_{AB}dx^Adx^B \\
	& +2 \rho^2 \theta^{(2)}_z dzdv + 2 \rho^2 \theta^{(2)}_{\bar{z}}  d\bar{z}dv +\rho^2 \lambda^{(2)}_{AB}dx^Adx^B +O(\rho^{2+\epsilon}),
	\end{split}
	\end{equation}
	where the first line  comes from a general spherically symmetric black hole solution like (\ref{Schwarzschild}), while the second and third lines are higher order terms which also contribute to the Einstein equations at the leading order. In matrix form it can be written as  
	\begin{equation}
	g_{\mu\nu}= \begin{pmatrix}
	-2\kappa \rho +(\frac{\rho}{r_h})^2+\rho^2a^{(2)} &1& \rho\theta_{A}+\rho^2 \theta^{(2)}_A\\
	1&0&0\\
	\rho\theta_{A}+\rho^2 \theta^{(2)}_A&0& (\rho+r_h)^2\gamma_{AB}+\rho\lambda_{AB}+\rho^2\lambda^{(2)}_{AB}\
	\end{pmatrix} +O(\rho^{2+\epsilon}).
	\end{equation} 
	Note that the conventions here are a little different from those in \cite{Donnay:2015abr} where their $\lambda_{AB}$ is the same as ours 
	$\tilde{\lambda}_{AB} = \lambda_{AB}+2r_h \gamma_{AB}$.

	\subsection{ Supertranslation and charge}

	The horizon supertranslation Killing vector~\footnote{As the above, we here also don't contain the vector associated with superrotation.} that preserves the asymptotic condition can be derived as \cite{Donnay:2016ejv} 
	\begin{equation}
	\xi=f(z,\bar{z})\partial_v +\(D_Af\int^{\rho}_0d\rho^{\prime}g^{AB}g_{vB}\) \partial
	_{\rho} - \(D_Bf\int^{\rho}_0d\rho^{\prime} g^{AB}\) \partial_A
	,
	\end{equation}
	or asymptotically\cite{Donnay:2015abr}
	\begin{equation}\label{supertranslation}
	\xi=f(z,\bar{z})\partial_v + \(\frac{\rho^2}{2r_h^2}\theta_AD^Af\) \partial_{\rho} +\(-\frac{\rho}{r_h^2}D^Af+\frac{\rho^2}{2r_h^4}
	{ \tilde{\lambda}^{AB} }
	D_B\) \partial_A + O(\rho^{3}),   
	\end{equation}
	which can generate the infinitesimal transformation    
	\begin{equation}\label{transformation}
	\begin{split}
	\mathcal{L}_{\xi}\theta_z&= -2\kappa D_zf(z,\bar{z}),\qquad \mathcal{L}_{\xi}\theta_{\bar{z}}= -2\kappa D_{\bar{z}}f(z,\bar{z}),\\
	\mathcal{L}_{\xi}\lambda_{AB}&= \theta_AD_Bf(z,\bar{z}) +\theta_B D_Af(z,\bar{z})-2D_AD_Bf(z,\bar{z}).
	\end{split}
	\end{equation}     
	For a general Schwarzschild black hole, one can get a more precise form \cite{Averin:2016ybl} 
	\begin{equation}
	\xi =f(z,\bar{z})\partial_v + D^z f(\frac{1}{r_h +\rho}-\frac{1}{r_h})\partial_z +  D^{\bar{z}}f(\frac{1}{r_h+\rho}-\frac{1}{r_h})\partial_{\bar{z}},
	\end{equation}
	which can lead to an  infinitesimal change given by Lie derivative 
	\begin{equation}\label{Lie}
	\mathcal{L}_{\xi}g_{\mu\nu}= 2\nabla_{(\mu}\xi_{\nu)}=\begin{pmatrix}
	0 &0& -\rho\frac{D_zf}{r} &-\rho\frac{D_{\bar{z}}f}{r}\\
	0&0&0&0\\
	-\rho\frac{D_zf}{r}&0&-\rho\frac{2rD_zD_zf}{r_h} &-\rho\frac{2rD_zD_{\bar{z}}f}{r_h}\\
	-\rho\frac{D_{\bar{z}}f}{r}&0&-\rho\frac{2rD_zD_{\bar{z}}f}{r_h} &-\rho\frac{2rD_{\bar{z}}D_{\bar{z}}f}{r_h}\\
	\end{pmatrix},
	\end{equation}
	where we have used $r=r_h +\rho$. The result has also been studied in \cite{Averin:2016ybl,Averin:2016hhm} where the authors discussed the interesting connection between Goldstone mode and quantum criticality \cite{Dvali:2012en}. 
	With the covariant approach developed in \cite{Barnich:2001jy,Barnich:2003xg}, one can get the charge related to the asymptotic Killing vector (\ref{supertranslation}) by calculating the variation of surface charge
	\begin{equation}\label{horizon charge}
	\begin{split}
	Q(f)& =\frac{2}{8 \pi G} \int_H dzd\bar{z} \gamma_{z\bar{z}} \Omega \kappa f(z,\bar{z})\\
	&= \frac{1}{4 \pi G} \int_H dzd\bar{z} \gamma_{z\bar{z}} m f(z,\bar{z}),
	\end{split}	
	\end{equation}
	where $H$ represents horizon and we have used $\kappa =1/2r_h,r_h=2m$ for a Schwarzschild black hole. In addition, we also added an extra factor 2 in the charge compared with the original definition in \cite{Donnay:2015abr,Donnay:2016ejv}. Note that they read the charge from the variation of it rather than directly calculating it. The factor 2 can be understood from the difference between the first law of black hole $\delta M = T\delta S$ and the Smarr formula $M =2TS$ in four dimensions. 
	Obviously, the extra factor 2 should be reasonable once  the fact  is considered that the charge should agree with the result of Komar integral which can be interpreted as the total energy of a stationary spacetime if we set $f(z,\bar{z})$ as 1.  
	According to the  standard definition, the Komar integral can be written as
	\begin{equation}
	E_t=\frac{1}{4\pi G} \int_{\partial \Sigma} d^2x \sqrt{\gamma^{(2)}}n_\mu\sigma_\nu \nabla^\mu K^\nu=M,
	\end{equation}
	where $\Sigma$ is a spacelike hypersurface and $ K^\mu=(1,0,0,0) $ is  Killing vector related to the time translation.

	Note that  the form of charge for horizon supertranslation is also the same as the one for supertranslation in BMS group. This feature should be related to the fact  that the ADM and Komar masses agree for stationary solution of general relativity \cite{Iyer:1994ys,Ashtekar:Abhay}. Of course, we can also use the method in \cite{Iyer:1994ys,Wald:1999wa} to calculate the Noether charge $(D-2)$ form of any infinitesimal diffeomorphism for any covariant gravity theory. In general relativity \footnote{It is worth checking whether all results about supertranslation and soft graviton still hold for any covariant gravity theory. } with or without cosmological constant, one can get the same result as the one in \cite{Donnay:2015abr}. From the relation between Noether charge and first law of black hole \cite{Iyer:1994ys,Wald:1999wa}, one can easily understand why the zero modes of charges defined in \cite{Donnay:2015abr,Donnay:2016ejv} correspond to entropy and angular momentum of stationary black hole. 	
	On the other hand,  the supertranslation charges commute with themselves \cite{Donnay:2015abr}. One can find that  the supertranslation (\ref{supertranslation}) will not change the energy of black hole because of the commutation 
	\begin{equation}
	\{Q(f),M\}=0.
	\end{equation}
	It will  not leave the black hole invariant but will only produce soft graviton as Goldstone bosons.  	

	\subsection{ Degenerate solutions}	
	All the results above do not depend on the equations of motion. Now we  consider the on-shell case in which  the Einstein equations without cosmological constant get satisfied\footnote{We put the analysis for the case of Einstein equations with a cosmological constant in Appendix \ref{AppB}.}
	\begin{equation}
	R_{\mu\nu}-\frac{1}{2}R g_{\mu\nu}= 8\pi G T_{\mu\nu}.
	\end{equation}
	In the vacuum case without any matter, they can be simplified as $R_{\mu\nu}=0$.
	Generally, there are ten components in metric $g_{\mu\nu}$ and ten Einstein equations, but only six of them are independent because of the four Bianchi identities. So we have the freedom to choose four coordinate conditions (\ref{gauge condition}). This looks like we can solve the whole Einstein equations and get some certain solutions. But as we have said before, there are still residual gauge invariances-supertranslations.  We can find how it can happen by solving the Einstein equations. We only consider the leading nontrivial order of the equations in what follows,  but of course we can solve them order by order and then get the complete solutions of the Einstein equations.
	First of all, by direct calculation of Ricci tensor one can find 
	\begin{equation}
	\lim_{\rho \rightarrow 0}R_{vv}=0,\quad \lim_{\rho \rightarrow 0}R_{vz}=-\frac{1}{2}\partial_v \theta_z=0  ,\quad  \lim_{\rho \rightarrow 0}R_{v\bar{z}}=-\frac{1}{2}\partial_v \theta_{\bar{z}}=0.
	\end{equation}
	So we can set $\theta_z$ and $\theta_{\bar{z}}$ as only functions of $(z,\bar{z})$. And one can get the other components of Ricci tensor at the leading order of $\rho$
	\begin{equation}\label{Ricci tensor}
	\begin{split}
	R_{\rho v} &= \frac{1}{2r_h^2}\left((2-4r_h \kappa)+ 2r_h^2a^{(2)} - 2\theta^z\theta_z -2\kappa {\lambda^z}_z +D^z\theta_z +D^{\bar{z}}\theta_{\bar{z}}-2\partial_v {\lambda^z}_z                             \right)\\ 	
	R_{\rho \rho} &= \frac{1}{2r_h^4}\left(4r_h {\lambda^{z}}_{z}  +\lambda^{\bar{z}z}\lambda_{z\bar{z}}+ \lambda^{zz}\lambda_{zz} -4r_h^2 {\lambda^{(2)z}}_z \right) \\	
	R_{\rho z} &= \frac{1}{2r_h^2}\left(2r_h^2\theta^{(2)}_z-\theta^z\lambda_{zz}+ D^z\lambda_{zz} -D^{\bar{z}}\lambda_{z\bar{z}} \right)     \\
	R_{\rho \bar{z}} &= \frac{1}{2r_h^2} \left(2r_h^2\theta^{(2)}_{\bar{z}}-\theta^{\bar{z}}\lambda_{\bar{z}\bar{z}}+ D^{\bar{z}}\lambda_{\bar{z}\bar{z}} -D^{z}\lambda_{z\bar{z}} \right)     \\
	R_{z\bar{z}} &=\frac{1}{2}\left(\gamma_{z\bar{z}}(2-4r_h\kappa)-\theta_z\theta_{\bar{z}} -2\kappa\lambda_{z\bar{z}}+D_{\bar{z}}\theta_z +D_z \theta_{\bar{z}}-2\partial_v\lambda_{z\bar{z}}                                   \right)        \\
	R_{zz} &=\frac{1}{2}\left(-\theta_z\theta_z -2\kappa\lambda_{zz} +D_z\theta_z +D_z \theta_z-2\partial_v\lambda_{zz}            \right)         \\                       
	R_{\bar{z}\bar{z}}&=\frac{1}{2}\left( -\theta_{\bar{z}}\theta_{\bar{z}} -2\kappa\lambda_{\bar{z}\bar{z}}+ D_{\bar{z}}\theta_{\bar{z}} +D_{\bar{z}} \theta_{\bar{z}}-2\partial_v\lambda_{\bar{z}\bar{z}}    \right)  ,       \\        
	\end{split}
	\end{equation}
	where we have taken the limit of $\rho \rightarrow 0$ and used  $\gamma^{z\bar{z}}$ to lift  the indices $z,\bar{z}$. Note that there are nine different functions that appear in the first order of all Ricci tensor components.  For the vacuum solution, we have  $R_{\mu\nu}=0$ . From $ R_{v\rho}=0=R_{z\bar{z}} $, one can get 
	\begin{equation}\label{solutions1}
	\begin{split}
	a^{(2)}&=\frac{\theta^z\theta_z}{2r_h^2}, \\
	\lambda_{z\bar{z}}& =r_h\left(D_{\bar{z}}\theta_z +D_z\theta_{\bar{z}}-\theta_z\theta_{\bar{z}}  + A e^{-\kappa v}     \right) .
	\end{split}
	\end{equation} 	 
	From other components in (\ref{Ricci tensor}), one can directly arrive at
	\begin{equation}\label{solutions2}
	\begin{split}
	\lambda_{zz}& =r_h\left(D_{z}\theta_z +D_z\theta_{z}-\theta_z\theta_{z}  + B e^{-\kappa v}     \right),\\
	\lambda_{\bar{z}\bar{z}}& =r_h\left(D_{\bar{z}}\theta_{\bar{z}} +D_{\bar{z}}\theta_{\bar{z}}-\theta_{\bar{z}}\theta_{\bar{z}}  + C e^{-\kappa v}     \right),\\
	\lambda^{(2)}_{z\bar{z}} &= \frac{1}{4r_h^2}\left( 4r_h\lambda_{z\bar{z}} +\lambda^z_z\lambda_{z\bar{z}} +\lambda^z_{\bar{z}}\lambda_{zz} \right),\\
	\theta^{(2)}_z &= \frac{1}{2r_h^2}\left( \theta^z\lambda_{zz}-D^z\lambda_{zz} +D^{\bar{z}}\lambda_{z\bar{z}}  \right),\\
	\theta^{(2)}_{\bar{z}} &= \frac{1}{2r_h^2}\left( \theta^{\bar{z}}\lambda_{\bar{z}\bar{z}}-D^{\bar{z}}\lambda_{\bar{z}\bar{z}} +D^{z}\lambda_{z\bar{z}}  \right).
	\end{split} 
	\end{equation}
	where we can represent them as the functions of $\theta_z$ and $\theta_{\bar{z}} $, and $A, B, C$ represent arbitrary functions of $(z,\bar{z})$, but are independent of $v$, which should be determined by the initial conditions.	 In appendix \ref{AppB}, we consider the Einstein equations with cosmological constant and get the solutions with the same forms as those in (\ref{solutions1}) and (\ref{solutions2}). The other components $R_{vv}$, $R_{vz}$, $R_{v\bar{z}}$ begin non-vanishing from the second order but also are not independent on others because of the Bianchi identities. On the contrary, we can use the those components to check the preceding solutions. For example, one can find 
	\begin{equation}
	\lim_{\rho \rightarrow 0} R_{vv}=- \frac{\rho}{2r_h^4} \left\lbrace   2r_h^3a^{(2)}  -2r_h\theta^z\theta_z -{\lambda^z}_z+r_h(D^z\theta_z +D^{\bar{z}}\theta_{\bar{z}}-\partial_v {\lambda^z}_z +2r_h \partial_v\partial_v {\lambda^z}_z)   \right\rbrace, 
	\end{equation}
	which is easy to show to be vanishing  when $\lambda_{z\bar{z}}$ and $a$ with the solution in (\ref{solutions1}) are substituted. Furthermore, we can also calculate $R_{vz}$ at the second order
	\begin{equation}
	\begin{split}
	\lim_{\rho \rightarrow 0} R_{vz}= - \frac{\rho}{4r_h^2}\big\lbrace &8r_h^2\kappa \theta^{(2)}_z +4\theta^z\theta_z\theta_z +2D_zD^z\theta_z -2D^{\bar{z}}D_z\theta_{\bar{z}} -4r_h^2D_za^{(2)} +4r_h^2\partial_v\theta^{(2)}_z\\
	&+\theta_z (8r_h \kappa +4\kappa{\lambda^z}_z -6D^z\theta_z +2D^{\bar{z}}\theta_{\bar{z}}+4\partial_v {\lambda^z}_z ) \\
	&+2\theta^z\partial_v\lambda_{zz}-2\partial_vD^z\lambda_{zz}+2\partial_vD^{\bar{z}}\lambda_{z\bar{z}}
	\big\rbrace,
	\end{split}	
	\end{equation}
	which is also equal to zero when we substitute $\lambda_{AB}$ and $\theta^{(2)}_z$ with the solution in (\ref{solutions2}) and notice that  $[D_z,D_{\bar{z}}]\theta_z=-\gamma_{z\bar{z}}\theta_z$.
	So from the all first order Einstein equations, we can not fix the whole components of metric which can influence these first order equations. This feature can be traced back to the fact that there is still residual diffeomorphism-supertranslation corresponding to the asymptotic killing vector (\ref{supertranslation}). All these infinitely degenerate black hole solutions can be related with each other by the supertranslation which generates the infinitesimal transformation (\ref{transformation}). For example, assuming $\theta_A$ has an infinitesimal transformation $\delta \theta_A=-2\kappa \partial_A f(z,\bar{z})$, one can get the transformation of $\lambda_{AB}$ from the solution (\ref{solutions2})
	\begin{equation}
	\delta \lambda_{AB}= -2D_AD_Bf(z,\bar{z})+\theta_AD_Bf(z,\bar{z}) +\theta_B D_Af(z,\bar{z}),
	\end{equation}
	which is compatible with Lie derivative of $\lambda_{AB}$ in (\ref{Lie}). In addition, there are exponentially decay modes $e^{-\kappa v}$ in these solutions. It can be related to the extension of supertranslation in \cite{Donnay:2016ejv} where they extend the form of function to allow $e^{-\kappa v}X(z,\bar{z})$ which can generate another new supertranslation.      
	But all these solutions only satisfy the vacuum Einstein equations, which does not mean they are all physical vacua with absence of all matter and radiation including gravitational radiation. Similar with the case in null infinity where BMS vacuum is defined by the vanishing of Bondi news which means  there are no radiative modes, we can define the physical vacuum with black hole as 
	\begin{equation}
	\partial_v \lambda_{AB}=0 \rightarrow  A=B=C=0,
	\end{equation} 
	which can define a stationary spacetime without radiation going in or out. On the other hand, $e^{-\kappa v}$ represents a kind of decay behaviour and must approach zero with the time $v$ increasing. So in the late time $v\rightarrow \infty$, we can get a fully static solution and the solution will return back to the physical vacuum state as expected. Obviously all this kind of physical vacua with black hole can be derived by supertranslation from the Schwarzschild black hole (\ref{Schwarzschild}) and written as (\ref{metric}) with 
	metric functions 
	\begin{equation}
	\begin{split}
	a^{(2)} & \approx 0,\\
	\theta_A&\approx  -2\kappa D_A f,\\
	\lambda_{AB}&\approx  -(D_AD_Bf+D_BD_Af), \quad... 
	\end{split}
	\end{equation}  
	where we have used the ellipsis to represent those higher order terms. If we only consider the approximation up to the first order of asymptotic Killing vector, the new metric by supertranslation from the Schwarzschild vacuum can be defined as $g_{\mu\nu}^{\prime}={g_0}_{\mu\nu}+\mathcal{L}_{\xi}{g_0}_{\mu\nu}$, where Lie derivatives is given in (\ref{Lie}). This has been discussed in Refs \cite{Averin:2016ybl,Eling:2016xlx}. Apparently, the new metric $g^{\prime}_{\mu\nu}$ will not be able to describe the vacuum solution near the horizon  because it ignored all higher order terms of asymptotic Killing vector or function $f(z,\bar{z})$. So we use (\ref{solutions1}-\ref{solutions2}) to represent the black hole solutions and label the physical vacuum through function $\theta(z,\bar{z})$ which can be related to $\theta_A$ and $\lambda_{AB}$ by  
	\begin{equation}
	\begin{split}
	a^{(2)}(z,\bar{z})&= \frac{D^z\theta D_z\theta}{2r_h^2}, \\
	\theta_A(z,\bar{z})&=-D_A\theta(z,\bar{z}), \\
	\lambda_{AB}(z\bar{z})& =r_h\left(-2D_{A}D_B\theta + D_A\theta D_B\theta   \right).\\
	\end{split}
	\end{equation}  
	Actually it is also possible for $\theta_A$ to contain some higher orders of functions $\theta(z,\bar{z})\approx2\kappa f(z,\bar{z})$. We need finite transformation to make sure of this point. BMS vacuum is determined by physical argument, see, e.g. (2.35) of \cite{Strominger:2013jfa} . The quantum state with black hole can be expressed as 
	\begin{equation}
	|M,\theta(z,\bar{z})\rangle \qquad  {\rm or} \qquad |M,C_{ln}\rangle \quad \text{with} \quad -l<n<l, ~-\infty <l<\infty
	\end{equation}  
	where one can use spherically harmonic functions $Y_{ln}(\theta,\phi)$ as basis to expand function $\theta(z,\bar{z})$ with expansion coefficients $C_{ln}$. These infinitely degenerate physical vacua can be distinguished by soft gravitons which play the role as Goldstone bosons of breaking horizon supertranslation symmetry and make black hole possible to storage information about how the black hole was formed or the initial state. Furthermore, the gravitational memory effect near the horizon will make us to detect the variation between two different vacuum states with black hole, whose  counterpart generated by supertranslation in BMS group is first pointed out by Strominger and Zhiboedov \cite{Strominger:2014pwa}. It was further illustrated by HPS in \cite{Hawking:2016msc} that degenerate black hole with infinite soft hairs can open a window for the  information loss paradox.\\

	\section{Discussion}

	Here we would like to discuss a little about the memory effect near the horizon.
	Assume a certain process generating  a black hole state to another one
	\begin{equation}
	|M,C_{ln}\rangle  \xrightarrow{radiation}   	|M',C'_{ln}\rangle, 
	\end{equation}
	where the latter with different mass can be considered as a result of the former black hole  absorbing some matter or emitting some radiation. Although thermal Hawking radiation contains no information, but the whole  spacetime or black hole horizon actually has the ability to store information about matter. It means that they can have different quantum state or quantum number $C_{l,n}$,  namely the spacetime carries with different information. To be honest, we did not repeat the  discussions in \cite{Strominger:2014pwa} to show the exact form of variation between two vacuum metrics induced by radiation. In principle, we can use the Einstein equations with radiation term to relate the degenerate black hole solutions and show that the variation can encode the information about the energy momentum tensor of radiation. We also did not use the charge defined in (\ref*{horizon charge}) to represent how to create a soft graviton. Technically, we don't have similar constraint equation like (\ref*{constraint equation}) because of $\partial_u m=0$. Obviously, the technical problem derives from our primary ansatz for stationary black hole or constant surface gravity $\kappa$. Physically, we don't introduce news tensor like $N_{zz}$ in the Bondi coordinates (\ref{Bondi coordinates}), because we pay attention on the on-shell degenerate solutions without radiation in this note. News tensors terms in the BMS supertranslation charges also be viewed as Goldstone of broken BMS supertranslation. For horizon supertranslation, ingoing expansion may have similar role because of similar transformation form. On the hand, Because we want to add radiation to generate the transformation between degenerate black holes, it may make more sense for us to use non-stationary spacetime or apparent horizon to describe the process although the initial and final black holes should be stationary. For example, the Vaidya spacetime
	\begin{equation}
	\begin{split}
	ds^2 = - \(1-\frac{2m(v)}{r}\)dv^2 + 2dv dr +2 r^2\gamma\indices{_{z\bar{z}}}dzd\bar{z},\\
	\qquad \text{with} \qquad  \partial_v m(v)=4\pi r^2 T_{vv},
	\end{split}
	\end{equation}
	is the simplest one with apparent horizon. But this metric satisfies the  BMS gauge fixing conditions in (\ref{BMS gauge}) rather than the gauges at the horizon  (\ref*{gauge condition}). It is found that Vaidya spacetime admits no BMS supertranslation field \cite{Compere:2016hzt} due to spherical symmetry. We hope to extend our calculations about horizon supertranslation to apparent horizon in the next work. On the other hand, the discrepancy in these two kinds of gauge fixing conditions also makes us unable to relate the horizon supertranslation with the BMS supertranslation. This point disagrees with the discussion about quotient space $BMS^H/BMS^-$ in \cite{Averin:2016ybl} where they only considered the special Schwarzschild metric $g_{0\mu\nu}$.

	\appendix

	\section{Higher order Ricci tensor and solutions}\label{AppA}
	In the main content  we only consider the nontrivial  leading order of the  components of all Ricci tensor to get the vacuum degenerate solutions. Here we give  these subleading terms of the components of the Ricci tensor, from which we can  get the next order components of the metric tensor. The metric conventions with  higher order terms take the form  
	\begin{equation}
	g_{\mu\nu}= \begin{pmatrix}
	-\frac{\rho}{r_h} +(\frac{\rho}{r_h})^2+\rho^2a^{(2)} + \rho^3  a^{(3)} &1& \rho\theta_{A}+ \rho^2 \theta^{(2)}_A +\rho^3\theta^{(3)}_A\\
	1&0&0\\
	\rho\theta_{A}+ \rho^2 \theta^{(2)}_A+ \rho^3 \theta^{(3)}_A&0& (\rho+r_h)^2\gamma_{AB}+\rho\lambda_{AB}+\rho^2\lambda^{(2)}_{AB}+\rho^3\lambda^{(3)}_{AB}\
	\end{pmatrix} +O(\rho^{3+\epsilon}),
	\end{equation} 
	with Ricci tensor defined in the way as
	\begin{equation}
	R_{\mu\nu}=R^{(0)}_{\mu\nu} + \rho R^{(1)}_{\mu\nu} + O (\rho^{2}).
	\end{equation}
	Besides the leading order terms given above, here we list the other components of Ricci tensor at the subleading order :      
	\def\z{{\bar{z}}}	
	
	\begin{equation}
	\begin{split}
	R^{(1)}_{\rho\rho}=& \frac{ -6 }{r_h^2}( {\lambda^{(3)z}}_{z} - F_{\rho\rho}),\\
	F_{\rho\rho}=& \frac{-1}{6r_h^4}\big\lbrace
	%6 r_h^4 {\lambda^{(3)z}}_{z}
	- 8 r_h^3 {\lambda^{(2)z}}_{z}
	- 2 r_h^2 ( {\lambda^{(2)}_{zz}} \lambda^{zz} +{\lambda^{(2)}_{ \z\z}} \lambda^{\z \z} + 2 {\lambda^{(2) }_{z\z}} \lambda^{z\z} ) \\
	& + 8 r_h^2 {\lambda^{z}}_{z}+ 6 r_h ( {\lambda^{z}}_z {\lambda^\z}_\z+ {\lambda^{\z}}_z {\lambda^z}_\z  )
	+ {\lambda^{z}}_z {\lambda^{z}}_z {\lambda^z}_z+ 3 {\lambda^{z}}_z   {\lambda^z}_\z  {\lambda^{\z}}_z 
	\big\rbrace,
	\end{split}	
	\end{equation}

	\begin{equation}
	\begin{split}
	R^{(1)}_{\rho v} =& 3(a^{(3)} -F_{\rho v}),\\
	F_{\rho v}=& \frac{-1}{ 12 r_h^4 }\big\lbrace  8(1+r_h\kappa+a^{(2)} r_h^2) r_h
	- 12 r_h^2(\theta^{(2)}_z \theta^z+\theta^{(2)}_\z \theta^\z)+ 4r_h(2\theta_z \theta^z-{\lambda^{(2)z}}_z) \\
	&+4( \theta^\z \theta_z {\lambda^z}_\z+\theta^z \theta_z {\lambda^z}_z+\theta^z \theta_\z {\lambda^\z}_z) +4\kappa({\lambda^z}_z{\lambda^\z}_\z+{\lambda^z}_\z{\lambda^\z}_z) %+2 r_h^2 a(4 r_h +2 {\lambda^z}_z)
	\\&+4r_h^2 (D^z\theta^{(2)}_z +D^\z\theta^{(2)}_\z )-  4 r_h (D^z\theta_z +D^\z\theta_\z )
	+4(1+4r_h \kappa +ar_h^2){\lambda^z}_z
	\\& - 2 (\theta^z D^z\lambda_{zz}+\theta^\z D^\z\lambda_{\z\z}
	+{\lambda^z}_z D^\z\theta_\z + {\lambda^z}_\z D^\z\theta_z
	+{\lambda^z}_zD^z\theta_z +{\lambda^{\bar{z}}}_zD^z\theta_{\bar{z}})
	\\&+ \partial_v(-8 r_h^2 {\lambda^{(2)z}}_z + 12 r_h {\lambda^{z}}_z +3 {\lambda^{z}}_z {\lambda^{\z}}_\z + 3 {\lambda^{z}}_\z {\lambda^{\z}}_z )
	\big\rbrace,
	\end{split}	
	\end{equation}

	\begin{equation}
	\begin{split}
	R^{(1)}_{\rho z}=&  3( \theta^{(3)}_{ z}- F_{\rho z}), \\
	F_{\rho z} =&   \frac{-1}{12r_h^4}\big\lbrace
	- 4 r_h^2 \theta_{(2)}^{ z} \lambda_{ z z}
	+4 r_h^2(D^z\lambda^{(2)}_{z z}-D^\z\lambda^{(2)}_{z\z}) - 8 r_h(D^z\lambda_{z z}-D^\z\lambda_{z\z}) 	\\
	&-2\theta_ z (2 r_h^2+2 r_h^2{\lambda^{(2) z}}_z- {\lambda^z}_\z {\lambda^\z}_z)
	+ \theta^z (- 8 r_h^2 \lambda^{(2)}_{zz} +3 (4 r_h + 2 {\lambda^{ z}}_z) {\lambda}_{z z}) \\
	&+({\lambda^{z}}_{\z}D^{\z} \lambda_{zz} +{\lambda^{\z}}_{z}D^{\z} \lambda_{\z\z}
	-2 {\lambda^{\z}}_{z}D^{z} \lambda_{z\z} - 4{\lambda^{z}}_{z}D^{z} \lambda_{z z} + 4 {\lambda^{z}}_{z}D^{\z} \lambda_{z\z}  )
	\big\rbrace,
	\end{split}	
	\end{equation}
	
	\begin{equation}
	\begin{split}
	R^{(1)}_{\rho \z}=&  3( \theta^{(3)}_{ z}- F_{\rho \z}), \\
	F_{\rho \z} =&   \frac{-1}{12r_h^4}\big\lbrace %12 r_h^4 \theta^{(3)}_{\z} 
	- 4 r_h^2 \theta_{(2)}^{\z} \lambda_{\z\z}
	-4 r_h^2(D^z\lambda^{(2)}_{z\z}-D^\z\lambda^{(2)}_{\z\z}) + 8 r_h(D^z\lambda_{z\z}-D^\z\lambda_{\z\z}) 	\\
	&-2\theta_\z (2 r_h^2+2 r_h^2{\lambda^{(2) z}}_z- {\lambda^z}_\z {\lambda^\z}_z)
	+ \theta^\z (- 8 r_h^2 \lambda^{(2)}_{\z\z} +3 (4 r_h + 2 {\lambda^{z}}_z) {\lambda}_{\z\z}) \\
	&+({\lambda^{z}}_{\z}D^{z} \lambda_{zz} +{\lambda^{\z}}_{z}D^{z} \lambda_{\z\z}
	-2 {\lambda^{z}}_{\z}D^{\z} \lambda_{z\z} - 4{\lambda^{z}}_{z}D^{\z} \lambda_{\z\z} + 4 {\lambda^{z}}_{z}D^{z} \lambda_{z\z}  )
	\big\rbrace,
	\end{split}	
	\end{equation}
	%	and
	~~
	
	\begin{equation}
	\begin{split}
	R^{(1)}_{z z} =& - 2(2\kappa +\partial_v )\lambda^{(2)}_{zz}+ 2F_{zz} ,\\
	F_{zz}  =&  \frac{1}{4r_h^2}\big\lbrace
	4 r_h^2 D_{z}\theta^{(2)}_{z}- 4 r_h^2 \theta_{z}\theta_{ z}^{(2)}  %- 4 r_h \lambda^{(2)}_{ z z}
	+2(2+2r_h\kappa+ r_h^2 a^{(2)} -\theta_z\theta^z) \lambda_{ z z} 
	+ 4 r_h \theta_{ z}\theta_{ z} \\
	&+ 2 \kappa (\lambda_{ z z} {\lambda^z}_{z} )
	+2 {\lambda^{z}}_{z} {\theta_{z}}\theta_{z}  
	+ \lambda_{ z z} D^{\z}\theta_{\z}-\lambda_{ z z} D^{z}\theta_{z}\\
	&	{-} \theta_{z} (4D^{\z}\lambda_{z\z} -3 D^{\z}\lambda_{ z z} )
	-  \theta_{\z}(D^{\z}\lambda_{ z z} )+\partial_v (2 r_h\lambda_{ z z} +  \lambda^{z}_{z}\lambda_{ z z})
	\big\rbrace,
	\end{split}	
	\end{equation}
	
	\begin{equation}
	\begin{split}
	R^{(1)}_{\z\z} =& - 2(2\kappa +\partial_v )\lambda^{(2)}_{\z\z}+ 2F_{\z\z} ,\\
	F_{\z\z} =&   \frac{1}{4r_h^2}\big\lbrace
	4 r_h^2 D_{\z}\theta^{(2)}_{\z}- 4 r_h^2 \theta_{\z}\theta_{\z}^{(2)} 
	+2(2+2r_h\kappa+ r_h^2 a^{(2)}-\theta_z\theta^z) \lambda_{\z\z} 
	+4 r_h \theta_{\z}\theta_{\z} \\
	&+ 2\kappa (\lambda_{\z \z} {\lambda^z}_{z} )
	+2 {\lambda^{z}}_{z} {\theta_{\z}}\theta_{\z} 
	+ \lambda_{\z\z} D^{z}\theta_z - \lambda_{\z\z} D^{\z}\theta_{\z} \\
	&	- \theta_{\z} (4D^{z}\lambda_{z\z} -3 D^{\z}\lambda_{\z\z} )
	-   \theta_{z}(D^{\z}\lambda_{\z\z} )+\partial_v(2 r_h\lambda_{\z\z} +  \lambda^{z}_{z}\lambda_{\z\z})
	\big\rbrace ,
	\end{split}	
	\end{equation}
	
	\begin{equation}
	\begin{split}
	R^{(1)}_{z\z}  =& \frac{1}{4r_h^4}\big\lbrace
	8r_h^3\gamma_{z\z} a^{(2)} + { 8 r_h D_{(z}\theta_{\z)} }
	- 8 r_h^2 \theta_{(z}\theta_{\z)}^{(2)} - 8 r_h \theta_{(z}\theta_{\z)}
	+4(1+ r_h^2 a^{(2)}-\theta_z\theta^z) \lambda_{z\z} \\
	& - 16 r_h^2\kappa \lambda^{(2)}_{z\z}+2\kappa(\lambda_{z z} {\lambda^z}_{\z} +\lambda_{\z \z} {\lambda^{\z}}_{z} )+2(\lambda_{z z} {\theta^z}\theta_{\z} +\lambda_{\z \z} \theta^{\z}\theta_{z})+2(\lambda_{z\z} D_{\z}\theta^\z+\lambda_{z\z} D_{z}\theta^{z})
	\\
	&	-2(\lambda_{\z\z} D_{z}\theta^\z+ \lambda_{z z} D_{\z}\theta^{z})
	+2(\theta^{\z}D_{\z}\lambda_{z\z} +\theta^{z}D_{z}\lambda_{z\z} )
	-4 (\theta^{\z}D_{z}\lambda_{\z\z} +\theta^{z}D_{\z}\lambda_{z z} )\\
	& + 8 r_h^2 D_{(z}\theta^{(2)}_{\z)} +2(D_{\z}D^{z}\lambda_{z z}-2 D_{z}D^{z}\lambda_{z \z}+D_{z}D^{\z}\lambda_{\z \z} )
	+\partial_v (-8r_h^2\lambda_{z\z}^{(2)} +2 {\lambda^{z}}_{\z}\lambda_{zz}) \big\rbrace.
	%	\\= \rho &\frac{1}{4r_h^2}\big\lbrace 	 4 r_h (2D_{(z}\theta_{\z)} -   \theta_{(z}\theta_{\z)} )	+\partial_t(-8r_h^2\lambda_{z\z}^{(2)} +2 \lambda^{z}_{\z}\lambda_{zz})  \\
	%  	&	+2 \lambda_{z\z} (4+ D_{\z}\theta^\z+  D_{z}\theta^{z} -\theta_z \theta^z)	- 8 r_h \lambda^{(2)}_{z\z} +\frac{2}{r_h}\lambda_{zz} {\lambda^z}_{\z} 	  	\big\rbrace 
	\end{split}	
	\end{equation}
	Note that  $R^{(1)}_{z\bar{z}}$ term is equal to zero if we consider the  solution of $\lambda^{(2)}_{z\bar{z}}$, so it is a trivial equation that can not help us to get higher order terms of the metric. But all other Ricci tensor components are enough for us to get the full vacuum solutions at the next order, although they can not be presented in a simple form. Note that we want to solve the equations $R^{(1)}_{\mu\nu}=0$ to get $\lambda^{(2)}_{AB}$, $a^{(3)}$, $\theta^{(2)}_A$, and it is easy to find that we firstly need $\lambda^{(2)}_{AB}$ except for  $\lambda^{(2)}_{z\bar{z}}$ that we have presented in (\ref{solutions1}).
	Actually from $R^{(1)}_{zz}=0$ or  $R^{(1)}_{\bar{z}\bar{z}}=0$, one can read a special kind of partial differential equations    
	\begin{equation}
	(2\kappa +\partial_v )\lambda^{(2)}_{AB}=F_{AB}(v,z,\bar{z}),
	\end{equation}
	where $F_{AB}(v,z,\bar{z})$ are completely determined by the lower order terms that we have been listed in $R^{(1)}_{AB}$ with a little complicate form. So the expected results take the form as 
	\begin{equation}
	\lambda^{(2)}_{AB} = D e^{-2\kappa v} +e^{-2\kappa v}\int^v_1  e^{2\kappa v'} F_{AB}(v',z,\bar{z})dv'   ,
	\end{equation} 
	where  $D$ is an arbitrary function of ($z,\bar{z})$.
	Then we can directly read off other components $\theta^{(3)}_A$, $a^{(3)}$ and $\lambda^{(3)}_{z\bar{z}}$ from $R^{(1)}_{\rho A}=0$, $R^{(1)}_{\rho v}=0$ and $R^{(1)}_{\rho \rho}=0$, respectively. 
	We do not  present all these components here, while  in principle higher order terms can also be solved order by order.

	\section{Degenerate (A)dS black hole} \label{AppB}
	
	In this appendix, we will present  the degenerate black hole solution  in (A)dS spacetime whose  Einstein equations contain a cosmological constant. Let us start with the (A)dS-Schwarzschild black hole   
	\begin{equation}
	ds^2 = - (1-\frac{2m}{r} - {\ll}\frac{r^2}{l^2})dv^2 + 2dv dr +2 r^2\gamma\indices{_{z\bar{z}}}dzd\bar{z},
	\end{equation}
	where $l$ represents the radius of (A)dS spacetime with ${\ll}= (-)1$ . The event horizon is determined by the equation 
	\begin{equation}
	1-\frac{2m}{r_h} -{\ll} \frac{r_h^2}{l^2}=0, \qquad \text{with\  surface\ gravity} \ \  \kappa=\frac{l^2-3{\ll} r_h^2}{2l^2r_h}.
	\end{equation}
	By introducing  a new radical coordinate $\rho
	=r-r_h$, one can arrive at 
	\begin{equation}
	\begin{split}
	ds^2 &=  \left[\left( -\frac{2m}{r_h^2}+\frac{2{\ll} r_h}{l^2}     \right)\rho +\left(\frac{{\ll}}{l^2}+\frac{2m}{r_h^3}\right)\rho^2          \right]dv^2 + 2dv d\rho +2 ( r_h +\rho)^2\gamma\indices{_{z\bar{z}}}dzd\bar{z} +O(\rho^{2+\epsilon}) \\
	&=\left(-2\kappa \rho +\frac{\rho^2}{r_h^2} \right)dv^2 + 2dvd\rho +2(\rho +r_h)^2 \gamma\indices{_{z\bar{z}}}dzd\bar{z}+O(\rho^{2+\epsilon}) ,
	\end{split}
	\end{equation}
	near the horizon,  whose difference from the leading order of  the Schwarzschild black hole solution is just the expression of surface gravity $\kappa$. So we can take the degenerate (A)dS-Schwarzschild black hole having the same asymptotic form as that in  (\ref{metric}), 
	\begin{equation}
	g_{\mu\nu}= \begin{pmatrix}
	-2\kappa \rho + \frac{\rho^2}{r_h^2}+\rho^2a^{(2)} &1& \rho\theta_{A}+ \rho^2 \theta^{(2)}_A\\
	1&0&0\\
	\rho\theta_{A}+ \rho^2 \theta^{(2)}_A &0& (\rho+r_h)^2\gamma_{AB}+\rho\lambda_{AB}+\rho^2\lambda^{(2)}_{AB}\
	\end{pmatrix} +O(\rho^{2+\epsilon}),
	\end{equation}       
	where only the  surface gravity is different from the one for the Schwarzschild black hole case.  
	As we said before, the definition of supertranslation is off-shell, and determined by the gauge conditions and asymptotic conditions. As a result,  one can find the same horizon supertranslation (\ref{supertranslation}) for the (A)dS-Schwarzschild black hole. This feature can be considered as an advantage of horizon supertranslation, compared with BMS supertranslation which is based on the null infinity of asymptotically flat spacetime, while  the null infinity is absent in asymptotically (A)dS spacetime.
	Next we consider the degenerate on-shell solutions which satisfy the Einstein equations with a cosmological constant 
	\begin{equation}
	R_{\mu\nu}-\frac{1}{2}g_{\mu\nu}R+\Lambda g_{\mu\nu}=0 \qquad \text{or} \qquad R_{\mu\nu}=\frac{2}{3}\Lambda g_{\mu\nu}=\ll\frac{2}{l^2}g_{\mu\nu}.
	\end{equation}  	
	Note that in this case, only the equations of $R_{vv}$, $R_{z\bar{z}}$ and $R_{vA}$ change at the leading order, compared to the case without the cosmological constant. From the results of (\ref{Ricci tensor}), one can find that  the solution read 
	\begin{equation}    
	\begin{split}
	a^{(2)} &=\frac{\theta^z\theta_z }{2r_h^2}, \\
	\lambda_{z\bar{z}}& =\frac{1}{2\kappa}\left( \gamma_{z\bar{z}}\frac{6r_h^2}{l^2}+\gamma_{z\bar{z}}(2-4\kappa r_h) +D_{\bar{z}}\theta_z +D_z\theta_{\bar{z}}-\theta_z\theta_{\bar{z}}  + A e^{-\kappa v}     \right)\\
	&= \frac{1}{2\kappa}\left(D_{\bar{z}}\theta_z +D_z\theta_{\bar{z}}-\theta_z\theta_{\bar{z}}  + A e^{-\kappa v}     \right),
	\end{split}
	\end{equation} 	
	and 
	\begin{equation}    
	\begin{split}
	\lambda_{zz}& =\frac{1}{2\kappa}\left(D_{z}\theta_z +D_z\theta_{z}-\theta_z\theta_{z}  + B e^{-\kappa v}     \right),\\
	\lambda_{\bar{z}\bar{z}}& =\frac{1}{2\kappa}\left(D_{\bar{z}}\theta_{\bar{z}} +D_{\bar{z}}\theta_{\bar{z}}-\theta_{\bar{z}}\theta_{\bar{z}}  + C e^{-\kappa v}     \right),\\
	\lambda^{(2)}_{z\bar{z}} &= \frac{1}{4r_h^2}\left( 4r_h\lambda_{z\bar{z}} +\lambda^z_z\lambda_{z\bar{z}} +\lambda^z_{\bar{z}}\lambda_{zz} \right),\\
	\theta^{(2)}_z &= \frac{1}{2r_h^2}\left( \theta^z\lambda_{zz}-D^z\lambda_{zz} +D^{\bar{z}}\lambda_{z\bar{z}}  \right),\\
	\theta^{(2)}_{\bar{z}} &= \frac{1}{2r_h^2}\left( \theta^{\bar{z}}\lambda_{\bar{z}\bar{z}}-D^{\bar{z}}\lambda_{\bar{z}\bar{z}} +D^{z}\lambda_{z\bar{z}}  \right).	
	\end{split}
	\end{equation} 	 	
	where we have used the definition of horizon to simplify the expression of $\lambda_{z\bar{z}}$. It is easy to find all results are totally equal to the solutions (\ref{solutions1}) and (\ref{solutions2}) for the  Schwarzschild black hole  case if one represents $\kappa$ by $1/2r_h$. Thus all discussions about infinitesimally degenerate black hole solutions keep valid for the (A)dS black hole as well.  	
	
	\section{Degenerate Rindler horizon }	\label{AppC}

	In this appendix, we discuss the Rindler horizon case  which is also studied in \cite{Hotta:2016qtv}. Just like the above, we can easily transform traditional Rindler metric into a new set of coordinates which satisfies the horizon gauge fixing conditions (\ref{gauge condition}). Starting from the standard Minkowski coordinates, one can arrive at    	 
	\begin{equation}
	\begin{split}
	ds^2 &=-dT^2+dX^2+dY^2+dZ^2  \\
	&=e^{2\kappa x}(-dt^2+dx^2) +dY^2+dZ^2,
	\end{split}
	\end{equation}	
	by the transformation between the inertial coordinate system and that of uniformly accelerated observer with acceleration $\kappa$ 	
	\begin{equation}
	X=\kappa^{-1}e^{\kappa x}\cosh\kappa t, \ \ \ T=\kappa^{-1}e^{\kappa x}\sinh\kappa t.
	\end{equation}
	Introducing  a new frame defined as 
	\begin{equation}
	e^{2\kappa x}=(1+\kappa \tilde{x})^2=2\kappa\rho, \ \ z=\frac{Y+iZ}{\sqrt{2}},\ \ \bar{z}=\frac{Y-iZ}{\sqrt{2}},
	\end{equation}
	we can rewrite the Rindler frame in the form 
	\begin{equation}
	\begin{split}
	ds^2 &=-(1+\kappa  \tilde{x})^2dt^2+d{\tilde{x}}^2+dY^2+dZ^2  \\
	&=-2\kappa\rho dt^2 +\frac{d\rho^2}{2\kappa\rho}+2dzd\bar{z},
	\end{split}
	\end{equation}	
	in which the horizon is located at $\rho=0$. This kind of  coordinates can also be obtained from a general black hole solution.  Finally, we can rewrite it to  the desired form  	
	\begin{equation}
	ds^2=-2\kappa\rho dv^2 +2dvd\rho +2dzd\bar{z},
	\end{equation}	
	with the transformation 
	\begin{equation}
	t \to v-g(\rho), \quad \rho \to e^{2\kappa g}.
	\end{equation} 	
	Consider degenerate solution in the near horizon region, we take the same asymptotic conditions with (\ref{asymptotic conditions}), and write the metric in the matrix form as  	
	\begin{equation}
	g_{\mu\nu}= \begin{pmatrix}
	-2\kappa \rho + \rho^2a^{(2)} &1& \rho\theta_{A}+ \rho^2 \theta^{(2)}_A\\
	1&0&0\\
	\rho\theta_{A}+ \rho^2\theta^{(2)}_A &0& \delta_{AB}+\rho\lambda_{AB}+\rho^2\lambda^{(2)}_{AB}\
	\end{pmatrix} +O(\rho^{2+\epsilon}),
	\end{equation}      	
	which corresponds to $\Omega(z,\bar{z})=\gamma^{z\bar{z}}$. From the results in \cite{Donnay:2015abr}, one can see that the horizon supertranslation for  the Rindler spacetime reads
	\begin{equation}\label{Rindler supertranslation}
	\xi=f(z,\bar{z})\partial_v +\(\frac{\rho^2\gamma_{z\bar{z}}}{2}\theta_AD^Af \)\partial_{\rho} +\left(-\rho\gamma_{z\bar{z}}D^Af+\frac{\rho^2\gamma_{z\bar{z}}^2}{2}\lambda^{AB}D_Bf\right)\partial_A + O(\rho^{3}),    
	\end{equation}	
	with conserved charge at horizon 	
	\begin{equation}
	Q(f) =\frac{1}{4 \pi G} \int_H dzd\bar{z}  \kappa f(z,\bar{z}).
	\end{equation}	
	Considering the vacuum Einstein equations $R_{\mu\nu}=0$ at the leading order in $\rho$, 
	one will arrive at the solution as follows,	
	\begin{equation}    
	\begin{split}
	a^{(2)} &=\frac{\theta_{\bar{z}}\theta_z }{2}, \\
	\lambda_{z\bar{z}}
	&= \frac{1}{2\kappa}\left(\partial_{\bar{z}}\theta_z +\partial_z\theta_{\bar{z}}-\theta_z\theta_{\bar{z}}  + A e^{-\kappa v}     \right),\\
	\lambda_{zz}& =\frac{1}{2\kappa}\left(\partial_{z}\theta_z +\partial_z\theta_{z}-\theta_z\theta_{z}  + B e^{-\kappa v}     \right),\\
	\lambda_{\bar{z}\bar{z}}& =\frac{1}{2\kappa}\left(\partial_{\bar{z}}\theta_{\bar{z}} +\partial_{\bar{z}}\theta_{\bar{z}}-\theta_{\bar{z}}\theta_{\bar{z}}  + C e^{-\kappa v}     \right),\\
	\lambda^{(2)}_{z\bar{z}} &= \frac{1}{4}\left(  \lambda_{z\bar{z}}\lambda_{z\bar{z}} +\lambda_{\bar{z}\bar{z}}\lambda_{zz} \right),\\
	\theta^{(2)}_z &= \frac{1}{2}\left( \theta_{\bar{z}}\lambda_{zz}-\partial_{\bar{z}}\lambda_{zz} +\partial_{z}\lambda_{z\bar{z}}  \right),\\
	\theta^{(2)}_{\bar{z}} &= \frac{1}{2}\left( \theta_{z}\lambda_{\bar{z}\bar{z}}-\partial_{z}\lambda_{\bar{z}\bar{z}} +\partial_{\bar{z}}\lambda_{z\bar{z}}  \right).
	\end{split}
	\end{equation} 	 	
	Here we did not give  explicit expressions for all components of Ricci tensor. But in order to check the preceding solution, we consider those  components which have not been used to get  the degenerate solution 
	\begin{equation}\label{Rindler solution}
	\begin{split}
	\lim_{\rho \rightarrow 0}R_{v\rho} &=\frac{ \rho}{2}\left(
	2a^{(2)}-2\theta_z\theta_{\bar{z}}-2\kappa\lambda_{z\bar{z}}+\partial_z\theta_{\bar{z}}+\partial_{\bar{z}}\theta_z-2\partial_v\lambda_{z\bar{z}}
	\right),\\
	\lim_{\rho \rightarrow 0}R_{vz}&= \frac{\rho}{2}\big(
	-4\kappa \theta^{(2)}_z -2\theta_{\bar{z}}\theta_z\theta_z -2\kappa \theta_z\lambda_{z\bar{z}} +3\theta_z\partial_{\bar{z}}\theta_z -\theta_z\partial_z\theta_{\bar{z}}-\partial_z\partial_{\bar{z}}\theta_z+\partial_z\partial_{\bar{z}}\theta_{\bar{z}} 
	\\
	&\qquad -2\partial_v\theta^{(2)}_z +2\partial_z a-2\theta_z\partial_v\lambda_{z\bar{z}}-\theta_{\bar{z}}\partial_v\lambda_{zz} +\partial_v\partial_{\bar{z}} \lambda_{zz}-\partial_v\partial_z \lambda_{z\bar{z}}
	\big).
	\end{split}
	\end{equation}   
	It is easy to show that they go to zero once  $\theta_A^{(2)}$ and $\lambda_{AB}$ are substituted  into (\ref{Rindler solution}). 
	All these degenerate Rindler solutions are physically distinguishable due to the soft gravitons, but are related to each other by the horizon supertranslation (\ref{Rindler supertranslation}). They can be considered as spacetime with the same kind of horizon because of the same acceleration or surface gravity $\kappa$, but with different information.

\acknowledgments

This work was finalized during a visit by R.G. Cai  as a visiting professor to the Yukawa Institute for Theoretical Physics, Kyoto University, the warm hospitality extended to him is greatly appreciated. We thank Pujian Mao's suggestions and discussions about the manuscript.
This work was supported in part by the  National Natural Science Foundation of China under Grants No.11375247 and No.11435006, and in part by a key project of CAS, Grant No.QYZDJ-SSW-SYS006.   	

\bibliography{biography_soft_hair}
\bibliographystyle{utphys}

\end{document}